\documentclass[aps,pra,english,twocolumn,showpacs,preprintnumbers,amsmath,amssymb,floatfix,superscriptaddress]{revtex4-1}
\usepackage[T1]{fontenc}
\usepackage[latin9]{inputenc}
\usepackage{graphicx}
\usepackage{amssymb}
\usepackage{babel}
\makeatletter

\usepackage[version=3]{mhchem}

\usepackage{color}

\usepackage{ulem}

\makeatother
\usepackage{babel}
\makeatother

\begin{document}

\title{Core-level spectra from graphene}

\author{Bo E. Sernelius}

\email{bos@ifm.liu.se}
\affiliation{Division of Theory and Modeling, Department of Physics, Chemistry and Biology, Link\"{o}ping University, SE-581 83 Link\"{o}ping, Sweden}

\begin{abstract}
We calculate core-level spectra for pristine and doped free-standing graphene sheets. Instructions for how to perform the calculations are given in detail. Although pristine graphene is not metallic the core-level spectrum presents low-energy tailing which is characteristic of metallic systems. The peak shapes vary with doping level in a characteristic way. The spectra are compared to experiments and show good agreement. We compare to two different pristine samples and to one doped sample. The pristine samples are one with quasi-free-standing epitaxial graphene on SiC obtained by hydrogen intercalation and one with a suspended graphene sheet. The doped sample is a gold supported graphene sheet. The gold substrate acts as an acceptor so the graphene sheet gets p-doped.
 \end{abstract}

\maketitle

\section{\label{introduction}Introduction}

In the 1970s many-body theory flourished. A topic that attracted much attention was the many-body effects on XPS (x-ray photoelectron spectroscopy) spectra, from deep core levels in metals\,\cite{Lang1,Lang2,Gad,Citrin1,Citrin2,DonSun}. Related effects were singularities that appeared near edges in absorption and emission spectra\,\cite{Mahan}. 

The cb (conduction-band) electrons can have many different effects on the XPS peak from the core level. If the excitation could be considered adiabatic the screening of the core hole by the cb  electrons would lead to a shift of the peak to higher kinetic energy (lower binding energy of the core level). Now the excitation is not adiabatic, the excitation is swift. The excitation frequency is large compared to the frequency components taking part in the screening of the core hole. This has the effect that the cb electrons are shaken up and are left in a nonequilibrium state. This is analogous to the following case: Assume that a cork is floating in a glass of water. If we remove the cork very slowly the water will be left very quiet and without any ripples. If we on the other hand remove it very briskly the water will be left in an upset state with many surface waves excited. In a metal the water waves correspond to plasmons, collective excitations of the cb electrons. 

The shake-up effects will show up as plasmon replicas in the spectrum; the main peak will be at the adiabatic position; the first replica, a smaller copy of the main peak, will be at a shifted position to lower kinetic energy where the shift is equal to the plasma frequency; the second replica is even smaller and shifted with two plasma frequencies; this goes on until further peaks are absorbed by the background. The main peak corresponds to the system left in quasi equilibrium(fully equilibrium would demand that the core hole were filled by one of the electrons), the second to a state where one plasmon is excited, the third to a state where two plasmons are excited and so on. Since all plasmons do not have identical energies, the plasmon curve shows dispersion, the replicas are not completely identical smaller versions of the main peak.

In a metal also single-particle excitations can take place. These electron-hole pair excitations form a continuum, starting from zero frequency and upwards. These excitations lead to a deformation, including a low-energy tail, of both the main peak and the plasmon replicas. 
Furthermore the finite life-time of the core hole causes a Lorentzian broadening of all peaks and experimental uncertainties give a Gaussian broadening.

The low-energy tailing is a characteristic of a metallic system, i.e. a system where the chemical potential is inside an energy band and not in a band-gap,  and can e.g. be used by experimentalists to find out where in a complex sample the core hole is situated. Several parameters were introduced by Doniach and \u{S}unji\'{c}\,\cite{DonSun} to characterize the XPS line shapes and are still broadly used by experimentalists in the fitting procedure for XPS spectra.

In the present work we address pristine and doped graphene. In the pristine case the chemical potential is neither inside an energy band nor in a band-gap. The Fermi surface is just two points in the Brillouin zone. This makes this system special. As we will see there is still a low-energy tailing. In the doped case the chemical potential is inside an energy band and we would expect to find, and find, a tailing. However, the 2D (two-dimensional) character of the system means that the collective excitations are 2D plasmons with a completely different dispersion than in the ordinary 3D metallic systems. The 2D plasmons give contributions that start already from zero frequency and upwards. This means that they contribute to the tail and no distinct plasmon replicas are distinguishable. Making any quantitative interpretation of graphene core-hole spectra using Doniach and \u{S}unji\'{c} fitting of the spectra is not feasible. It is more reasonable to calculate the spectra. One purpose of this work is to provide the reader with the tools needed for such a calculation. 

The material is organized in the following way: In Sec. \ref{the} we show how the core-hole spectra are derived for a 2D system. The results and comparison with experimental spectra are presented in Sec. \ref{res}. Finally, we end with a brief summary and conclusion section, Sec. \ref{sum}.

\section{\label{the}Theory}
We use a model that is based on the one used by Langreth\,\cite{Lang} for the core-hole problem in the 1970s and here modified and extended to fit our problem. We have earlier\,\cite{SerSod} with success used another modified version for the problem of exciton annihilation in quantum wells. 

In the excitation process the photoelectron leaves the system and a core hole is left behind. The shape of the XPS spectrum depends on how fast the process is. If it is very slow one may use the adiabatic approximation in which one assumes that the electrons in the system have time to relax around the core hole during the process. 

When we derive the XPS line shape we assume that the excitation process is very fast; we use the sudden approximation in which the core-hole potential is turned on instantaneously.  The electrons have not time,  during the process, to settle down and reach equilibrium in the potential from the core hole. This results in shake-up effects in the form of single particle (electron-hole pair) excitations and collective (plasmon) excitations. The electrons contributing to the shake-up effects are the electrons in the valence and conduction bands. From here on we refer to them as the electrons.

We use the assumption that the core hole does not recoil in the shake-up process and that there are no excitations within the core.  
We approximate the core-hole potential with a pure Coulomb potential.

The following model Hamiltonian for the system is used:
\begin{equation}
H = {\varepsilon}C_{ch}^\dag {C_{ch}} + C_{ch}^\dag {C_{ch}}V + {H_e},
\label{equ1}
\end{equation}
where ${\varepsilon }$, $C_{ch}^\dag $, ${C_{ch}}$, $V$, and ${H_e}$ are the core-hole energy, the creation operator for the core hole, the annihilation operator for the core hole, the interaction potential for the interaction between the core hole and the electrons, and the Hamiltonian for the electrons, respectively. The operators $V$ and ${H_e}$ contain creation and annihilation operators for the electrons and no core-hole operators. 

The core-hole number operator is ${n_{ch}} = C_{ch}^\dag {C_{ch}}$. It commutes with the Hamiltonian, i.e., $\left[ {{n_{ch}},H} \right] = 0$.

In the ground state of the system there is no core hole and the electrons are in their ground state. Let us introduce two Hamiltonians
\begin{equation}
\begin{array}{l}
{H^{\left( 0 \right)}} = {H_e},\\
{H^{\left( 1 \right)}} = \varepsilon  + V + {H_e},
\end{array}
\label{equ2}
\end{equation}
where the first is the Hamiltonian before the core hole has been created and the second the Hamiltonian after the creation.
Let $\left| 0 \right\rangle  = {\left| 0 \right\rangle _e}{\left| 0 \right\rangle _{ch}}$ denote the ground state of $H$. It is also the ground state of ${H^{\left( 0 \right)}}$. Then we have
\begin{equation}
\begin{array}{l}
{H_e}\left| 0 \right\rangle  = E_0^e\left| 0 \right\rangle ,\\
H\left| 0 \right\rangle  = E_0^e\left| 0 \right\rangle ,\\
{H^{\left( 0 \right)}}\left| 0 \right\rangle  = E_0^e\left| 0 \right\rangle ,
\end{array}
\label{equ3}
\end{equation}
where  $E_0^e$ is the ground state energy of the electron system. From now on we drop the subscript $ch$ on the core-hole operators and
introduce the functions
\begin{equation}
\left\{ \begin{array}{l}
{G^ > }(t) = \left\langle 0 \right|C\left( t \right){C^\dag }\left( 0 \right)\left| 0 \right\rangle \\
{G^ < }(t) = \left\langle 0 \right|{C^\dag }\left( 0 \right)C\left( t \right)\left| 0 \right\rangle 
\end{array} \right..
\label{equ4}
\end{equation}

These functions are connected to the time ordered and retarded Green's functions, according to
 \begin{equation}
{G^T}(t) =  - i\left\langle 0 \right|TC\left( t \right){C^\dag }\left( 0 \right)\left| 0 \right\rangle  = \left\{ \begin{array}{l}
 - i{G^ > }(t)\,\,;\,\,t > 0\\
i{G^ < }(t)\,\,;\,\,t < 0,
\end{array} \right.
\label{equ5}
\end{equation}
and
\begin{equation}
\begin{array}{l}
{G^R}(t) =  - i\theta \left( t \right)\left\langle 0 \right|\left[ {C\left( t \right),{C^\dag }\left( 0 \right)} \right]\left| 0 \right\rangle \\
\quad \quad  =  - i\theta \left( t \right)\left[ {{G^ > }(t) - {G^ < }(t)} \right].
\end{array}
\label{equ6}
\end{equation}

The Fourier transformed versions,
\begin{equation}
{G^{\scriptstyle < \hfill\atop
\scriptstyle > \hfill}}\left( \omega  \right) = \int\limits_{ - \infty }^\infty  {dt{e^{-i\omega t}}{G^{\scriptstyle < \hfill\atop
\scriptstyle > \hfill}}\left( t \right)}, 
\label{equ7}
\end{equation}
have a direct physical meaning. Their sum is the spectral function, ${G^ < }\left( \omega  \right)/2\pi $ is the density of states for an occupied core-hole state, and ${G^ > }\left( \omega  \right)/2\pi $ for an unoccupied. Note that the creation and annihilation operators are for holes. We want the density of states for a core electron. Thus we need to calculate ${G^ > }\left( \omega  \right)/2\pi $.

 We first determine the time dependent form. We have
 \begin{equation}
\begin{array}{l}
{G^ > }(t) = \left\langle 0 \right|C\left( t \right){C^\dag }\left( 0 \right)\left| 0 \right\rangle \\
 = \left\langle 0 \right|{e^{iHt/\hbar }}C{e^{ - iHt/\hbar }}{C^\dag }\left| 0 \right\rangle \\
 = {e^{iE_0^et/\hbar }}\left\langle 0 \right|C{e^{ - iHt/\hbar }}{C^\dag }\left| 0 \right\rangle \\
 = {e^{iE_0^et/\hbar }}\left\langle 0 \right|C{e^{ - i\left[ {\varepsilon n + nV + {H_e}} \right]t/\hbar }}{C^\dag }\left| 0 \right\rangle \\
 = {e^{iE_0^et/\hbar }}\left\langle 0 \right|C{e^{ - i\varepsilon nt/\hbar }}{e^{ - i\left[ {nV + {H_e}} \right]t/\hbar }}{C^\dag }\left| 0 \right\rangle \\
 = {e^{i\left( {E_0^e - \varepsilon } \right)t/\hbar }}\left\langle 0 \right|C{e^{ - i\left[ {V + {H_e}} \right]t/\hbar }}{C^\dag }\left| 0 \right\rangle \\
 = {e^{i\left( {E_0^e - \varepsilon } \right)t/\hbar }}{}_e\left\langle 0 \right|{e^{ - i\left[ {V + {H_e}} \right]t/\hbar }}{\left| 0 \right\rangle _e}{}_{ch}\left\langle 0 \right|1 - n{\left| 0 \right\rangle _{ch}}\\
 = {e^{i\left( {E_0^e - \varepsilon } \right)t/\hbar }}\left\langle 0 \right|{e^{ - i\left[ {V + {H_e}} \right]t/\hbar }}\left| 0 \right\rangle ,
\end{array}
\label{equ8}
\end{equation}
where we have let the operators operate to the left and right. We have made use of the general relation ${e^{A + B}} = {e^A}{e^B}{e^{ - {\textstyle{1 \over 2}}\left[ {A,B} \right]}}$, for operators $A$ and $B$, and that the core-hole number operator commutes with all terms of the Hamiltonian. It is now time to define our $V$ and the rest of the Hamiltonian. 

The excitation spectrum of the electrons is given by the dynamical structure factor, which is related to the dielectric function according to
\begin{equation}
S({\bf{q}},\omega ) =  - \frac{{\hbar  }}{{{v_q}}}{\mathop{\rm Im}\nolimits} {\varepsilon ^{ - 1}}({\bf{q}},\omega )
\label{equ9}
\end{equation}
where ${v_q} = 2\pi {e^2}/q$ is the 2D Fourier transform of the Coulomb potential. The structure factor is nonzero in two regions; one region is where electron-hole pairs are excited; the other is on the plasmon dispersion curve. All these excitations are bosons. We assume that the bosons are independent. 

Let us for simplicity start by assuming that the bosons have a distinct dispersion curve, $\omega \left( {\bf{q}} \right) = {\omega _{\bf{q}}}$. This is the case for plasmons and phonons. Later we will generalize this to include contributions from the electron-hole pair continuum. We now get the following Hamiltonian:
\begin{equation}
\begin{array}{l}
H' = V + {H_e}\\
 =  - \frac{1}{{{A^{{1 \mathord{\left/
 {\vphantom {1 2}} \right.
 \kern-\nulldelimiterspace} 2}}}}}\sum\limits_{\bf{q}} {g({\bf{q}})\rho _{ch}^\dag ({\bf{q}})} ({c_{\bf{q}}} + c_{ - {\bf{q}}}^\dag ) + \sum\limits_{\bf{q}} {\hbar {\omega _{\bf{q}}}(c_{\bf{q}}^\dag {c_{\bf{q}}} + {1 \mathord{\left/
 {\vphantom {1 2}} \right.
 \kern-\nulldelimiterspace} 2}),} 
\end{array}
\label{equ10}
\end{equation}
where $A$ is the area of the 2D system and ${g({\bf{q}})}$ the coupling constant between the core hole and the boson excitation. The operators ${c_{\bf{q}}^\dag }$ and  ${{c_{\bf{q}}}}$ are boson creation and annihilation operators, respectively, obeying the commutation relation $\left[ {{c_{\bf{q}}},c_{{\bf{q}}'}^\dag } \right] = {\delta _{{\bf{q}},{\bf{q}}'}}$ and
\begin{equation}
\begin{array}{c}
\rho _{ch}^\dag ({\bf{q}}) = \int {{d^2}r} {e^{i{\bf{q}} \cdot {\bf{r}}}}\delta \left( {{\bf{r}} - {{\bf{R}}_{ch}}} \right) = {e^{i{\bf{q}} \cdot {{\bf{R}}_{ch}}}},\\
{\rho _{ch}}({\bf{q}}) = \int {{d^2}r} {e^{ - i{\bf{q}} \cdot {\bf{r}}}}\delta \left( {{\bf{r}} - {{\bf{R}}_{ch}}} \right) = {e^{ - i{\bf{q}} \cdot {{\bf{R}}_{ch}}}},
\end{array}
\label{equ11}
\end{equation}
are the density operators, where ${{\bf{R}}_{ch}}$ denotes the position of the core hole.
We treat the core hole as a classical particle, i.e., the core-hole operators are c-numbers. This is why the density operators do not contain any creation and/or annihilation operators. 

Let us here take the opportunity to derive the results using the adiabatic approximation. Since there is no kinetic energy terms for the core hole the Hamiltonian can be diagonalized. This is achieved by using the following unitary transformation:
\begin{equation}
U = \exp \left[ {\sum\limits_{\bf{q}} {{f^\dag }({\bf{q}})} ({c_{\bf{q}}} - c_{ - {\bf{q}}}^\dag )} \right],
\label{equ12}
\end{equation}
where
\begin{equation}
{f^\dag }({\bf{q}}) = f( - {\bf{q}}) = \frac{{g({\bf{q}})}}{{{A^{{1 \mathord{\left/
 {\vphantom {1 2}} \right.
 \kern-\nulldelimiterspace} 2}}}\hbar {\omega _{\bf{q}}}}}\rho _{ch}^\dag ({\bf{q}}).
\label{equ13}
\end{equation}
Letting
\begin{equation}
B = \sum\limits_{\bf{q}} {{f^\dag }({\bf{q}})} ({c_{\bf{q}}} - c_{ - {\bf{q}}}^\dag ),
\label{equ14}
\end{equation}
we find that $\left[ {B,{c_{\bf{q}}}} \right] = {f^\dag }( - {\bf{q}}) = f({\bf{q}})$ and  $\left[ {B,{c^\dag }_{\bf{q}}} \right] = {f^\dag }({\bf{q}})$.
This gives
\begin{equation}
\begin{array}{l}
U{c_{\bf{q}}}{U^\dag } = {c_{\bf{q}}} + f({\bf{q}}),\\
Uc_{\bf{q}}^\dag {U^\dag } = c_{\bf{q}}^\dag  + {f^\dag }({\bf{q}}).
\end{array}
\label{equ15}
\end{equation}

The transformation of the Hamiltonian gives
\begin{equation}
\begin{array}{l}
UH'{U^\dag } = \sum\limits_{\bf{q}} {\hbar {\omega _{\bf{q}}}\left[ {\left( {c_{\bf{q}}^\dag  + {f^\dag }({\bf{q}})} \right)\left( {{c_{\bf{q}}} + f({\bf{q}})} \right) + {1 \mathord{\left/
 {\vphantom {1 2}} \right.
 \kern-\nulldelimiterspace} 2}} \right]} \\
 - \frac{1}{{{A^{{1 \mathord{\left/
 {\vphantom {1 2}} \right.
 \kern-\nulldelimiterspace} 2}}}}}\sum\limits_{\bf{q}} {g({\bf{q}})\rho _{ch}^\dag ({\bf{q}})} \left( {{c_{\bf{q}}} + f({\bf{q}}) + c_{ - {\bf{q}}}^\dag  + {f^\dag }( - {\bf{q}})} \right)\\
 = \sum\limits_{\bf{q}} {\hbar {\omega _{\bf{q}}}\left( {c_{\bf{q}}^\dag {c_{\bf{q}}} + {1 \mathord{\left/
 {\vphantom {1 2}} \right.
 \kern-\nulldelimiterspace} 2}} \right)}  - \frac{1}{{{A^{{1 \mathord{\left/
 {\vphantom {1 2}} \right.
 \kern-\nulldelimiterspace} 2}}}}}\sum\limits_{\bf{q}} {g({\bf{q}})\rho _{ch}^\dag ({\bf{q}})} \left( {{c_{\bf{q}}} + c_{ - {\bf{q}}}^\dag } \right)\\
 + \sum\limits_{\bf{q}} {\hbar {\omega _{\bf{q}}}\left( {c_{\bf{q}}^\dag f({\bf{q}}) + {c_{\bf{q}}}{f^\dag }({\bf{q}}) + {f^\dag }({\bf{q}})f({\bf{q}})} \right)} \\
 - \frac{2}{{{A^{{1 \mathord{\left/
 {\vphantom {1 2}} \right.
 \kern-\nulldelimiterspace} 2}}}}}\sum\limits_{\bf{q}} {g({\bf{q}})\rho _{ch}^\dag ({\bf{q}})} f({\bf{q}}).
\end{array}
\label{equ16}
\end{equation}

Substituting for the expression of $f({\bf{q}})$ we find
\begin{equation}
\begin{array}{l}
UH'{U^\dag } = \sum\limits_{\bf{q}} {\hbar {\omega _{\bf{q}}}\left( {c_{\bf{q}}^\dag {c_{\bf{q}}} + {1 \mathord{\left/
 {\vphantom {1 2}} \right.
 \kern-\nulldelimiterspace} 2}} \right)}  + \\
 - \frac{1}{{{A^{{1 \mathord{\left/
 {\vphantom {1 2}} \right.
 \kern-\nulldelimiterspace} 2}}}}}\sum\limits_{\bf{q}} {\left( {{c_{\bf{q}}} + c_{ - {\bf{q}}}^\dag } \right)\overbrace {\left( {g({\bf{q}})\rho _{ch}^\dag ({\bf{q}}) - {A^{{1 \mathord{\left/
 {\vphantom {1 2}} \right.
 \kern-\nulldelimiterspace} 2}}}\hbar {\omega _{\bf{q}}}{f^\dag }({\bf{q}})} \right)}^0} \\
 - \frac{1}{A}\sum\limits_{\bf{q}} {\frac{{{{\left| {g({\bf{q}})} \right|}^2}}}{{\hbar {\omega _{\bf{q}}}}}\rho _{ch}^\dag ({\bf{q}})} {\rho _{ch}}({\bf{q}}).
\end{array}
\label{equ17}
\end{equation}

Thus
 \begin{equation}
UH'{U^\dag } = \sum\limits_{\bf{q}} {\hbar {\omega _{\bf{q}}}\left( {c_{\bf{q}}^\dag {c_{\bf{q}}} + {1 \mathord{\left/
 {\vphantom {1 2}} \right.
 \kern-\nulldelimiterspace} 2}} \right)}  + \Delta \varepsilon,
\label{equ18}
\end{equation}

where
\begin{equation}
\begin{array}{l}
\Delta \varepsilon  =  - \frac{1}{A}\sum\limits_{\bf{q}} {\frac{{{{\left| {g({\bf{q}})} \right|}^2}}}{{\hbar {\omega _{\bf{q}}}}}\rho _{ch}^\dag ({\bf{q}})} {\rho _{ch}}({\bf{q}})\\
 =  - \frac{1}{A}\sum\limits_{\bf{q}} {\frac{{{{\left| {g({\bf{q}})} \right|}^2}}}{{\hbar {\omega _{\bf{q}}}}}{e^{i{\bf{q}} \cdot {{\bf{R}}_{ch}}}}{e^{ - i{\bf{q}} \cdot {{\bf{R}}_{ch}}}}}  =  - \frac{1}{A}\sum\limits_{\bf{q}} {\frac{{{{\left| {g({\bf{q}})} \right|}^2}}}{{\hbar {\omega _{\bf{q}}}}}}. 
\end{array}
\label{equ19}
\end{equation}

We see that the transformation of the Hamiltonian in Eq.\,(\ref{equ10}) has the effect that the interaction term is changed into a constant energy term. The interactions with the core hole produce an energy shift of the ground state. This is the relaxation energy; the gain in energy when the electrons relax around the core hole. This is the shift of the XPS peak (towards higher kinetic energy) one would get in the adiabatic approximation. 

We now return to the sudden approximation and our Green's function ${G^ > }(t) = {e^{i\left( {E_0^h - \varepsilon } \right){t \mathord{\left/
 {\vphantom {t \hbar }} \right.
 \kern-\nulldelimiterspace} \hbar }}}\left\langle 0 \right|{e^{ - i\left[ {H'} \right]{t \mathord{\left/
 {\vphantom {t \hbar }} \right.
 \kern-\nulldelimiterspace} \hbar }}}\left| 0 \right\rangle $. Let us insert the identity on both sides of the exponential inside the ground state matrix element.
 \begin{equation}
\begin{array}{l}
{G^ > }(t) = {e^{i\left( {E_0^h - \varepsilon } \right){t \mathord{\left/
 {\vphantom {t \hbar }} \right.
 \kern-\nulldelimiterspace} \hbar }}}\left\langle 0 \right|{U^\dag }U{e^{ - iH'{t \mathord{\left/
 {\vphantom {t \hbar }} \right.
 \kern-\nulldelimiterspace} \hbar }}}{U^\dag }U\left| 0 \right\rangle \\
 = {e^{i\left( {E_0^h - \varepsilon } \right){t \mathord{\left/
 {\vphantom {t \hbar }} \right.
 \kern-\nulldelimiterspace} \hbar }}}\left\langle 0 \right|{U^\dag }{e^{ - iUH'{U^\dag }{t \mathord{\left/
 {\vphantom {t \hbar }} \right.
 \kern-\nulldelimiterspace} \hbar }}}U\left| 0 \right\rangle \\
 = {e^{i\left( {E_0^h - \varepsilon } \right){t \mathord{\left/
 {\vphantom {t \hbar }} \right.
 \kern-\nulldelimiterspace} \hbar }}}\left\langle 0 \right|{U^\dag }{e^{ - i\left[ {\sum\limits_{\bf{q}} {\hbar {\omega _{\bf{q}}}\left( {c_{\bf{q}}^\dag {c_{\bf{q}}} + {1 \mathord{\left/
 {\vphantom {1 2}} \right.
 \kern-\nulldelimiterspace} 2}} \right)}  + \Delta \varepsilon } \right]{t \mathord{\left/
 {\vphantom {t \hbar }} \right.
 \kern-\nulldelimiterspace} \hbar }}}U\left| 0 \right\rangle. 
\end{array}
\label{equ20}
\end{equation}

Now, since there is no boson excited in the ground state we may write
\begin{equation}
\begin{array}{l}
{e^{i\left[ {\sum\limits_{\bf{q}} {\hbar {\omega _{\bf{q}}}\left( {c_{\bf{q}}^\dag {c_{\bf{q}}} + {1 \mathord{\left/
 {\vphantom {1 2}} \right.
 \kern-\nulldelimiterspace} 2}} \right)} } \right]{t \mathord{\left/
 {\vphantom {t \hbar }} \right.
 \kern-\nulldelimiterspace} \hbar }}}\left| 0 \right\rangle  = {e^{i\left( {\sum\limits_{\bf{q}} {{\textstyle{1 \over 2}}\hbar {\omega _{\bf{q}}}} } \right){t \mathord{\left/
 {\vphantom {t \hbar }} \right.
 \kern-\nulldelimiterspace} \hbar }}}\left| 0 \right\rangle \\
 = {e^{iE_0^h{t \mathord{\left/
 {\vphantom {t \hbar }} \right.
 \kern-\nulldelimiterspace} \hbar }}}\left| 0 \right\rangle.
\end{array}
\label{equ21}
\end{equation}
Thus we have
\begin{equation}
\begin{array}{l}
{G^ > }(t) = {e^{ - i\left( {\varepsilon  + \Delta \varepsilon } \right){t \mathord{\left/
 {\vphantom {t \hbar }} \right.
 \kern-\nulldelimiterspace} \hbar }}}\\
\times \left\langle 0 \right|{U^\dag }{e^{ - i\left[ {\sum\limits_{\bf{q}} {\hbar {\omega _{\bf{q}}}\left( {c_{\bf{q}}^\dag {c_{\bf{q}}} + {1 \mathord{\left/
 {\vphantom {1 2}} \right.
 \kern-\nulldelimiterspace} 2}} \right)} } \right]{t \mathord{\left/
 {\vphantom {t \hbar }} \right.
 \kern-\nulldelimiterspace} \hbar }}}U{e^{i\left[ {\sum\limits_{\bf{q}} {\hbar {\omega _{\bf{q}}}\left( {c_{\bf{q}}^\dag {c_{\bf{q}}} + {1 \mathord{\left/
 {\vphantom {1 2}} \right.
 \kern-\nulldelimiterspace} 2}} \right)} } \right]{t \mathord{\left/
 {\vphantom {t \hbar }} \right.
 \kern-\nulldelimiterspace} \hbar }}}\left| 0 \right\rangle \\
 = {e^{ - i\left( {\varepsilon  + \Delta \varepsilon } \right){t \mathord{\left/
 {\vphantom {t \hbar }} \right.
 \kern-\nulldelimiterspace} \hbar }}}\left\langle 0 \right|{U^\dag }\left( 0 \right)U\left( t \right)\left| 0 \right\rangle ,
\end{array}
\label{equ22}
\end{equation}
where
\begin{equation}
U\left( t \right) = uU{u^\dag };\;u = {e^{ - i\left( {\sum\limits_{\bf{q}} {{\omega _{\bf{q}}}c_{\bf{q}}^\dag {c_{\bf{q}}}} } \right)t}},
\label{equ23}
\end{equation}
and we see that $u$ is also a unitary transformation. Now, from Eqs.\,(\ref{equ12}) and (\ref{equ23}) we find
\begin{equation}
\begin{array}{l}
U(t) = \exp \left[ {\sum\limits_{\bf{q}} {{f^\dag }({\bf{q}})} (\underbrace {u{c_{\bf{q}}}{u^\dag }}_{{c_{\bf{q}}}{e^{i{\omega _{\bf{q}}}t}}} - \underbrace {uc_{ - {\bf{q}}}^\dag {u^\dag }}_{c_{ - {\bf{q}}}^\dag {e^{ - i{\omega _{\bf{q}}}t}}})} \right]\;\\
 = \exp \left[ {\sum\limits_{\bf{q}} {{f^\dag }({\bf{q}})} ({c_{\bf{q}}}{e^{i{\omega _{\bf{q}}}t}} - c_{ - {\bf{q}}}^\dag {e^{ - i{\omega _{\bf{q}}}t}})} \right]\\
 = \exp \left[ {\sum\limits_{\bf{q}} {{f^\dag }({\bf{q}})} ( - c_{ - {\bf{q}}}^\dag {e^{ - i{\omega _{\bf{q}}}t}} + {c_{\bf{q}}}{e^{i{\omega _{\bf{q}}}t}})} \right].
\end{array}
\label{equ24}
\end{equation}

If we now once again make use of the relation ${e^{A + B}} = {e^A}{e^B}{e^{ - {\textstyle{1 \over 2}}\left[ {A,B} \right]}}$ we find
\begin{equation}
\begin{array}{l}
U(t) = \exp \left[ { - \sum\limits_{\bf{q}} {{f^\dag }({\bf{q}})} c_{ - {\bf{q}}}^\dag {e^{ - i{\omega _{\bf{q}}}t}}} \right]\exp \left[ {\sum\limits_{\bf{q}} {{f^\dag }({\bf{q}})} {c_{\bf{q}}}{e^{i{\omega _{\bf{q}}}t}}} \right]\\
\quad \quad  \times \exp \left[ {{\textstyle{1 \over 2}}\sum\limits_{{\bf{k}},{\bf{l}}} {{f^\dag }({\bf{l}})} {f^\dag }({\bf{k}}){e^{ - i{\omega _{\bf{l}}}t}}{e^{i{\omega _{\bf{k}}}t}}\underbrace {\left[ {c_{ - {\bf{l}}}^\dag ,{c_{\bf{k}}}} \right]}_{ - {\delta _{{\bf{k}}, - {\bf{l}}}}}} \right]\\
 = \exp \left[ { - \sum\limits_{\bf{q}} {{f^\dag }({\bf{q}})} c_{ - {\bf{q}}}^\dag {e^{ - i{\omega _{\bf{q}}}t}}} \right]\exp \left[ {\sum\limits_{\bf{q}} {{f^\dag }({\bf{q}})} {c_{\bf{q}}}{e^{i{\omega _{\bf{q}}}t}}} \right]\\
\quad \quad  \times \exp \left[ { - {\textstyle{1 \over 2}}\sum\limits_{\bf{q}} {{f^\dag }({\bf{q}})} {f^\dag }( - {\bf{q}})} \right]\\
 = \exp \left[ { - \sum\limits_{\bf{q}} {{f^\dag }({\bf{q}})} c_{ - {\bf{q}}}^\dag {e^{ - i{\omega _{\bf{q}}}t}}} \right]\exp \left[ {\sum\limits_{\bf{q}} {{f^\dag }({\bf{q}})} {c_{\bf{q}}}{e^{i{\omega _{\bf{q}}}t}}} \right]\\
\quad \quad  \times \exp \left[ { - {\textstyle{1 \over 2}}\sum\limits_{\bf{q}} {{f^\dag }({\bf{q}})} f({\bf{q}})} \right]\\
 = \exp \left[ { - \sum\limits_{\bf{q}} {{f^\dag }({\bf{q}})} c_{ - {\bf{q}}}^\dag {e^{ - i{\omega _{\bf{q}}}t}}} \right]\exp \left[ {\sum\limits_{\bf{q}} {{f^\dag }({\bf{q}})} {c_{\bf{q}}}{e^{i{\omega _{\bf{q}}}t}}} \right]\\
\quad \quad  \times \exp \left[ { - {\textstyle{1 \over 2}}\sum\limits_{\bf{q}} {{{\left| {f({\bf{q}})} \right|}^2}} } \right].
\end{array}
\label{equ25}
\end{equation}

We also need
\begin{equation}
\begin{array}{l}
{U^\dag }(0) = \exp \left[ {\sum\limits_{\bf{q}} {f({\bf{q}})} c_{\bf{q}}^\dag } \right]\exp \left[ { - \sum\limits_{\bf{q}} {f({\bf{q}})} {c_{ - {\bf{q}}}}} \right]\\
\quad \quad  \times \exp \left[ { - {\textstyle{1 \over 2}}\sum\limits_{\bf{q}} {{{\left| {f({\bf{q}})} \right|}^2}} } \right].
\end{array}
\label{equ26}
\end{equation}

Substituting the results of  Eqs.\,(\ref{equ25}) and (\ref{equ26}) in Eq.\,(\ref{equ22}) gives
\begin{equation}
\begin{array}{l}
{G^ > }(t) = {e^{ - i\left( {\varepsilon  + \Delta \varepsilon } \right){t \mathord{\left/
 {\vphantom {t \hbar }} \right.
 \kern-\nulldelimiterspace} \hbar }}}\left\langle 0 \right|{U^\dag }\left( 0 \right)U\left( t \right)\left| 0 \right\rangle \\
 = {e^{ - i\left( {\varepsilon  + \Delta \varepsilon } \right){t \mathord{\left/
 {\vphantom {t \hbar }} \right.
 \kern-\nulldelimiterspace} \hbar }}}\left\langle 0 \right|\exp \left[ {\sum\limits_{\bf{q}} {f({\bf{q}})} c_{\bf{q}}^\dag } \right]\exp \left[ { - \sum\limits_{\bf{q}} {f({\bf{q}})} {c_{ - {\bf{q}}}}} \right]\\
\quad \quad  \times \exp \left[ { - {\textstyle{1 \over 2}}\sum\limits_{\bf{q}} {{{\left| {f({\bf{q}})} \right|}^2}} } \right]\exp \left[ { - \sum\limits_{\bf{q}} {{f^\dag }({\bf{q}})} c_{ - {\bf{q}}}^\dag {e^{ - i{\omega _{\bf{q}}}t}}} \right]\\
\quad \quad  \times \exp \left[ {\sum\limits_{\bf{q}} {{f^\dag }({\bf{q}})} {c_{\bf{q}}}{e^{i{\omega _{\bf{q}}}t}}} \right]\exp \left[ { - {\textstyle{1 \over 2}}\sum\limits_{\bf{q}} {{{\left| {f({\bf{q}})} \right|}^2}} } \right]\left| 0 \right\rangle.
\end{array}
\label{equ27}
\end{equation}

The factors that do not contain boson operators can be taken outside. Then we can let the left most factor operate to the left. It produces a factor of unity. The rightmost factor may operate to the right and it also produces a factor of unity. We are left with
\begin{equation}
\begin{array}{l}
{G^ > }(t) = {e^{ - i\left( {\varepsilon  + \Delta \varepsilon } \right){t \mathord{\left/
 {\vphantom {t \hbar }} \right.
 \kern-\nulldelimiterspace} \hbar }}}{e^{ - \sum\limits_{\bf{q}} {{{\left| {f({\bf{q}})} \right|}^2}} }}\\
\quad \quad  \times \left\langle 0 \right|{e^{ - \sum\limits_{\bf{q}} {f({\bf{q}})} {c_{ - {\bf{q}}}}}}{e^{ - \sum\limits_{\bf{q}} {{f^\dag }({\bf{q}})} c_{ - {\bf{q}}}^\dag {e^{ - i{\omega _{\bf{q}}}t}}}}\,\left| 0 \right\rangle. 
\end{array}
\label{equ28}
\end{equation}

Here we make use of the relation ${e^A}{e^B} = {e^B}{e^A}{e^{\left[ {A,B} \right]}}$ and find
\begin{equation}
\begin{array}{l}
{G^ > }(t) = \exp \left[ { - i\left( {\varepsilon  + \Delta \varepsilon } \right){t \mathord{\left/
 {\vphantom {t \hbar }} \right.
 \kern-\nulldelimiterspace} \hbar }} \right]\exp \left[ { - \sum\limits_{\bf{q}} {{{\left| {f({\bf{q}})} \right|}^2}} } \right]\\
 \times \left\langle 0 \right|\exp \left[ { - \sum\limits_{\bf{q}} {{f^\dag }({\bf{q}})} c_{ - {\bf{q}}}^\dag {e^{ - i{\omega _{\bf{q}}}t}}} \right]\\
 \times \exp \left[ { - \sum\limits_{\bf{q}} {f({\bf{q}})} {c_{ - {\bf{q}}}}} \right]\\
 \times \exp \left\{ {\left[ { - \sum\limits_{\bf{q}} {f({\bf{q}})} {c_{ - {\bf{q}}}}, - \sum\limits_{\bf{q}} {{f^\dag }({\bf{q}})} c_{ - {\bf{q}}}^\dag {e^{ - i{\omega _{\bf{q}}}t}}} \right]} \right\}\left| 0 \right\rangle \\
 = \exp \left[ { - i\left( {\varepsilon  + \Delta \varepsilon } \right){t \mathord{\left/
 {\vphantom {t \hbar }} \right.
 \kern-\nulldelimiterspace} \hbar }} \right]\exp \left[ { - \sum\limits_{\bf{q}} {{{\left| {f({\bf{q}})} \right|}^2}} } \right]\\
 \times \left\langle 0 \right|\exp \left[ { - \sum\limits_{\bf{q}} {f({\bf{q}})} {c_{ - {\bf{q}}}}} \right]\\
 \times \exp \left\{ {\left[ { - \sum\limits_{\bf{q}} {f({\bf{q}})} {c_{ - {\bf{q}}}}, - \sum\limits_{\bf{q}} {{f^\dag }({\bf{q}})} c_{ - {\bf{q}}}^\dag {e^{ - i{\omega _{\bf{q}}}t}}} \right]} \right\}\left| 0 \right\rangle \\
 = \exp \left[ { - i\left( {\varepsilon  + \Delta \varepsilon } \right){t \mathord{\left/
 {\vphantom {t \hbar }} \right.
 \kern-\nulldelimiterspace} \hbar }} \right]\exp \left[ { - \sum\limits_{\bf{q}} {{{\left| {f({\bf{q}})} \right|}^2}} } \right]\\
 \times \left\langle 0 \right|\exp \left[ { - \sum\limits_{\bf{q}} {f({\bf{q}})} {c_{ - {\bf{q}}}}} \right]\\
 \times \exp \left\{ {\sum\limits_{{\bf{k}},{\bf{l}}} {f({\bf{k}}){f^\dag }({\bf{l}}){e^{ - i{\omega _{\bf{l}}}t}}} \underbrace {\left[ {{c_{ - {\bf{k}}}},c_{ - {\bf{l}}}^\dag } \right]}_{{\delta _{{\bf{k}},{\bf{l}}}}}} \right\}\left| 0 \right\rangle \\
 = \exp \left[ { - i\left( {\varepsilon  + \Delta \varepsilon } \right){t \mathord{\left/
 {\vphantom {t \hbar }} \right.
 \kern-\nulldelimiterspace} \hbar }} \right]\exp \left[ { - \sum\limits_{\bf{q}} {{{\left| {f({\bf{q}})} \right|}^2}} } \right]\\
 \times \exp \left[ {\sum\limits_{\bf{q}} {{{\left| {f({\bf{q}})} \right|}^2}{e^{ - i{\omega _{\bf{q}}}t}}} } \right]\\
 \times \left\langle 0 \right|\exp \left[ { - \sum\limits_{\bf{q}} {f({\bf{q}})} {c_{ - {\bf{q}}}}} \right]\left| 0 \right\rangle \\
 = {e^{ - i\left( {\varepsilon  + \Delta \varepsilon } \right){t \mathord{\left/
 {\vphantom {t \hbar }} \right.
 \kern-\nulldelimiterspace} \hbar }}}{e^{ - \sum\limits_{\bf{q}} {{{\left| {f({\bf{q}})} \right|}^2}} }}{e^{\sum\limits_{\bf{q}} {{{\left| {f({\bf{q}})} \right|}^2}{e^{ - i{\omega _{\bf{q}}}t}}} }}.
\end{array}
\label{equ29}
\end{equation}

Thus we have
\begin{equation}
{G^ > }(t) = {e^{ - i\left( {{\varepsilon  \mathord{\left/
 {\vphantom {\varepsilon  \hbar }} \right.
 \kern-\nulldelimiterspace} \hbar } + \Delta {\varepsilon  \mathord{\left/
 {\vphantom {\varepsilon  \hbar }} \right.
 \kern-\nulldelimiterspace} \hbar }} \right)t}}{e^{ - \sum\limits_{\bf{q}} {{{\left| {f({\bf{q}})} \right|}^2}\left( {1 - {e^{ - i{\omega _{\bf{q}}}t}}} \right)} }}.
\label{equ30}
\end{equation}

We want the Fourier transformed version. It is
\begin{equation}
\begin{array}{l}
{G^ > }(\omega ) = \int\limits_{ - \infty }^\infty  {dt} {e^{ - i\omega t}}{G^ > }(t)\\
\quad \quad  = \int\limits_{ - \infty }^\infty  {dt} {e^{ - i\left( {\omega  + {\varepsilon  \mathord{\left/
 {\vphantom {\varepsilon  \hbar }} \right.
 \kern-\nulldelimiterspace} \hbar } + \Delta {\varepsilon  \mathord{\left/
 {\vphantom {\varepsilon  \hbar }} \right.
 \kern-\nulldelimiterspace} \hbar }} \right)t}}{e^{B\left( t \right)}},
\end{array}
\label{equ31}
\end{equation}
where the so-called satellite generator is
\begin{equation}
B\left( t \right) =  - \sum\limits_{\bf{q}} {{{\left| {f({\bf{q}})} \right|}^2}\left( {1 - {e^{ - i{\omega _{\bf{q}}}t}}} \right)}. 
\label{equ32}
\end{equation}
\begin{figure}
\includegraphics[width=6cm]{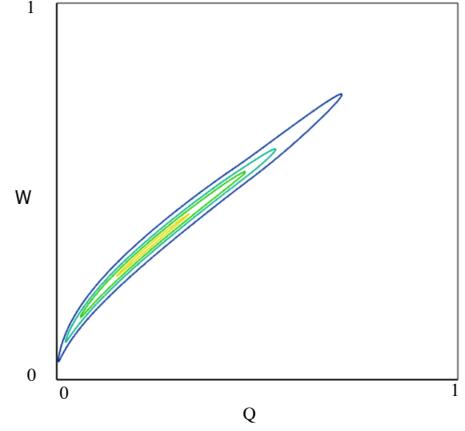}
\caption{(Color online) Contour plot of $-{\mathop{\rm Im}\nolimits} \left[ {\varepsilon {{\left( {Q,W} \right)}^{ - 1}}} \right]$ for doped graphene. We see the typical 2D plasmon dispersion. The plot is for the doping density $10^{14}  \rm{cm}^{-2}$ but the result is very similar for any finite doping density. }
\label{figu1}
\end{figure}

In the unperturbed case, i.e., when there is no interaction with the core hole we have
 \begin{equation}
G_0^ > (\omega ) = \int\limits_{ - \infty }^\infty  {dt} {e^{ - i\left( {\omega  + {\varepsilon  \mathord{\left/
 {\vphantom {\varepsilon  \hbar }} \right.
 \kern-\nulldelimiterspace} \hbar }} \right)t}} = \int\limits_{ - \infty }^\infty  {dt} {e^{ - i\omega t}}G_0^ > (t),
\label{equ33}
\end{equation}
so we can write
\begin{equation}
{G^ > }(\omega ) = \int\limits_{ - \infty }^\infty  {dt} {e^{ - i\omega t}}G_0^ > (t){e^{C\left( t \right)}},
\label{equ34}
\end{equation}
where
\begin{equation}
\begin{array}{l}
C\left( t \right) = B\left( t \right) - i\Delta \varepsilon {t \mathord{\left/
 {\vphantom {t \hbar }} \right.
 \kern-\nulldelimiterspace} \hbar }\\
\quad \quad  =  - \sum\limits_{\bf{q}} {{{\left| {f({\bf{q}})} \right|}^2}\left( {1 - i{\omega _{\bf{q}}}t - {e^{ - i{\omega _{\bf{q}}}t}}} \right)} \\
\quad \quad  =  - \frac{1}{A}\sum\limits_{\bf{q}} {\frac{{{{\left| {g({\bf{q}})} \right|}^2}{{\left| {{\rho _{ch}}({\bf{q}})} \right|}^2}}}{{{{\left( {\hbar {\omega _{\bf{q}}}} \right)}^2}}}\left( {1 - i{\omega _{\bf{q}}}t - {e^{ - i{\omega _{\bf{q}}}t}}} \right)} \\
\quad \quad  =  - \frac{1}{A}\sum\limits_{\bf{q}} {\frac{{{{\left| {g({\bf{q}})} \right|}^2}}}{{{{\left( {\hbar {\omega _{\bf{q}}}} \right)}^2}}}\left( {1 - i{\omega _{\bf{q}}}t - {e^{ - i{\omega _{\bf{q}}}t}}} \right)}.
\label{equ35}
\end{array}
\end{equation}
Now we generalize this result to include all excitation processes in our system. We first rewrite Eq. (\ref{equ35}) 
\begin{equation}
\begin{array}{l}
C\left( t \right) =  - \frac{1}{A}\sum\limits_{\bf{q}} {\frac{{{{\left| {g({\bf{q}})} \right|}^2}}}{{{{\left( {\hbar {\omega _{\bf{q}}}} \right)}^2}}}\left( {1 - i{\omega _{\bf{q}}}t - {e^{ - i{\omega _{\bf{q}}}t}}} \right)} \\
\quad \quad  =  - \frac{1}{A}\sum\limits_{\bf{q}} {\int {\frac{{d\omega }}{{2\pi }}} \underbrace {2\pi \delta \left( {\omega  - {\omega _{\bf{q}}}} \right)}_{ \equiv D_{\bf{q}}^ > \left( \omega  \right)}\frac{{{{\left| {g({\bf{q}})} \right|}^2}}}{{{{\left( {\hbar \omega } \right)}^2}}}} \\
\quad \quad \quad \quad \quad \quad  \times \left( {1 - i\omega t - {e^{ - i\omega t}}} \right)
\end{array}
\label{equ36}
\end{equation}
where ${D_{\bf{q}}^ > \left( \omega  \right)}$ is the boson propagator. Now the boson propagator is replaced by $2$ times the dynamical structure factor of Eq.\,(\ref{equ9}) and the coupling constant by ${v_q}$. Thus we get
\begin{equation}
\begin{array}{l}
C\left( t \right) =  - \frac{1}{A}\sum\limits_{\bf{q}} {\int {\frac{{d\omega }}{{2\pi }}} {{\left( {{v_q}} \right)}^2}2\pi \frac{\hbar }{{\pi {v_q}}}\left( { - {\mathop{\rm Im}\nolimits} \frac{1}{{\varepsilon \left( {{\bf{q}},\omega } \right)}}} \right)} \\
\quad \quad \quad \quad \quad \quad  \times \frac{1}{{{{\left( {\hbar \omega } \right)}^2}}}\left( {1 - i\omega t - {e^{ - i\omega t}}} \right)\\
 = \frac{2}{\pi }\int\limits_0^\infty  {\int\limits_0^\infty  {dqd\omega {e^2}} } \frac{1}{{\hbar {\omega ^2}}}{\mathop{\rm Im}\nolimits} \frac{1}{{\varepsilon \left( {q,\omega } \right)}}\left( {1 - i\omega t - {e^{ - i\omega t}}} \right).
\end{array}
\label{equ37}
\end{equation}
\begin{figure}
\includegraphics[width=5.5cm]{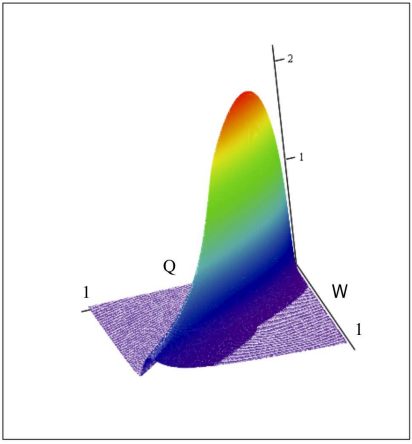}
\caption{(Color online) Surface plot of $-{\mathop{\rm Im}\nolimits} \left[ {\varepsilon {{\left( {Q,W} \right)}^{ - 1}}} \right]$ for doped graphene. The plot is for the doping density $10^{14}  \rm{cm}^{-2}$ but the result is very similar for any finite doping density.}
\label{figu2}
\end{figure}
The XPS spectrum reflects the density of states of the core electron and becomes
\begin{equation}
\begin{array}{l}
S\left( \omega  \right) = \frac{1}{{2\pi }}\int\limits_{ - \infty }^\infty  {dt{e^{ - i\omega t}}{e^{C\left( t \right)}}} \\
 = \frac{1}{{2\pi }}\int\limits_{ - \infty }^\infty  {dt{e^{ - i\left( {\omega  - d} \right)t}}{e^{ - a(t)}}{e^{ - ib\left( t \right)}}} \\
 = \frac{1}{{2\pi }}\int\limits_{ - \infty }^\infty  {dt{e^{ - i\left[ {\left( {\omega  - d} \right)t + b\left( t \right)} \right]}}{e^{ - a(t)}}} \\
 = \frac{1}{\pi }\int\limits_0^\infty  {dt\cos \left[ {\left( {\omega  - d} \right)t + b\left( t \right)} \right]{e^{ - a(t)}}}, 
\end{array}
\label{equ38}
\end{equation}
where
\begin{equation}
\begin{array}{l}
a(t) = \frac{{2{e^2}}}{{\pi \hbar }}\int\limits_0^\infty  {\int\limits_0^\infty  {dqd\omega } } \frac{1}{{{\omega ^2}}}{\mathop{\rm Im}\nolimits} \frac{{ - 1}}{{\varepsilon \left( {q,\omega } \right)}}\left[ {1 - \cos \left( {\omega t} \right)} \right],\\
b\left( t \right) =  - \frac{{2{e^2}}}{{\pi \hbar }}\int\limits_0^\infty  {\int\limits_0^\infty  {dqd\omega } } \frac{1}{{{\omega ^2}}}{\mathop{\rm Im}\nolimits} \frac{{ - 1}}{{\varepsilon \left( {q,\omega } \right)}}\sin \left( {\omega t} \right),\\
d = \frac{{2{e^2}}}{{\pi \hbar }}\int\limits_0^\infty  {\int\limits_0^\infty  {dqd\omega } } \frac{1}{\omega }{\mathop{\rm Im}\nolimits} \frac{{ - 1}}{{\varepsilon \left( {q,\omega } \right)}}.
\end{array}
\label{equ39}
\end{equation}
We give the energy relative to the core-level position when the interaction with the electrons are neglected.
Taking the finite life time of the core hole into account we arrive at the central theoretical result of this work, the XPS spectrum.

The XPS spectrum can be written as
\begin{equation}
S(W) = \frac{1}{\pi }\int\limits_0^\infty  {{e^{\left( { - \frac{{LW}}{2}T} \right)}}{e^{ - a(T)}}\cos \left[ {\left( {W - D} \right)T - b\left( T \right)} \right]dT}, 
\label{equ40}
\end{equation}
where
\begin{equation}
\begin{array}{l}
a(T)\\
 = \frac{{{e^2}{k_F}}}{{\pi {E_F}}}\int\limits_0^\infty  {\int\limits_0^\infty  {\frac{1}{{{W^2}}}{\rm{Im}}\left[ {\frac{{ - 1}}{{\varepsilon \left( {Q,W} \right)}}} \right]\left[ {1 - \cos \left( {WT} \right)} \right]dQ} dW} ,
\end{array}
\label{equ41}
\end{equation}
\begin{equation}
b(T) =  - \frac{{{e^2}{k_F}}}{{\pi {E_F}}}\int\limits_0^\infty  {\int\limits_0^\infty  {\frac{1}{{{W^2}}}{\rm{Im}}\left[ {\frac{{ - 1}}{{\varepsilon \left( {Q,W} \right)}}} \right]\sin \left( {WT} \right)dQ} dW},
\label{equ42}
\end{equation}
and
\begin{equation}
D = \frac{{{e^2}{k_F}}}{{\pi {E_F}}}\int\limits_0^\infty  {\int\limits_0^\infty  {\frac{1}{W}{\rm{Im}}\left[ {\frac{{ - 1}}{{\varepsilon \left( {Q,W} \right)}}} \right]dQ} dW}.
\label{equ43}
\end{equation}

\begin{figure}
\includegraphics[width=5.5cm]{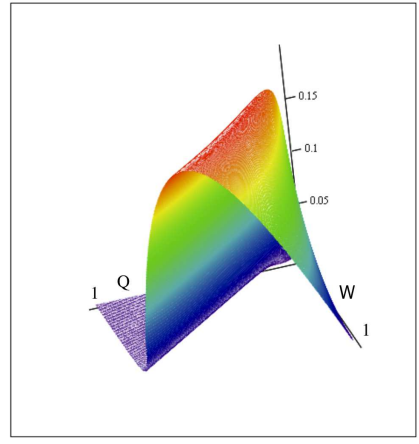}
\caption{(Color online) Surface plot of $-{\mathop{\rm Im}\nolimits} \left[ {\varepsilon {{\left( {Q,W} \right)}^{ - 1}}} \right]$ for pristine graphene. The scaling of $W$ and $Q$ is the same as in Fig.\,\ref{figu2}.}
\label{figu3}
\end{figure}

All variables have been scaled according to
\begin{equation}
\begin{array}{l}
Q = q/2{k_F};\\
W = \hbar \omega /2{E_F};\\
T = t2{E_F}/\hbar ;\\
D = d/2{E_F};\\
LW = lw/2{E_F};\\
GW = gw/2{E_F};\\
{W_0} = \hbar {\omega _0}/2{E_F},
\end{array}
\label{equ44}
\end{equation}
and are now dimensionless. The two last relations will be used below. The quantity $d$ is the energy shift of the adiabatic peak. We have taken the finite life-time of the core hole into account by introducing the first factor of the integrand of Eq.(\ref{equ40}), where ${lw}$ is the FWHM (full width at half maximum) of the Lorentz broadened peak. We have here assumed that the core-hole potential can be represented by a pure Coulomb potential. The results are valid for a general 2D system. The particular system enters the problem through ${\mathop{\rm Im}\nolimits} \left[ {\varepsilon {{\left( {Q,W} \right)}^{ - 1}}} \right]$. For pristine graphene the dielectric function is\,\cite{Guinea}
\begin{equation}
\varepsilon \left( {Q,W} \right) = \kappa  + \frac{{\pi {e^2}}}{{2\hbar v}}\frac{Q}{{\sqrt {{Q^2} - {W^2}} }},
\label{equ45}
\end{equation}
where we have used the values\,\cite{Wun}  $8.73723 \times {10^5}$ m/s for the Fermi velocity $v$ and $2.4$ for $\kappa$, respectively. The background dielectric constant $\kappa$ is the result of high frequency electronic excitations to higher lying empty energy bands and from lower lying filled energy bands.

\begin{figure}
\includegraphics[width=8cm]{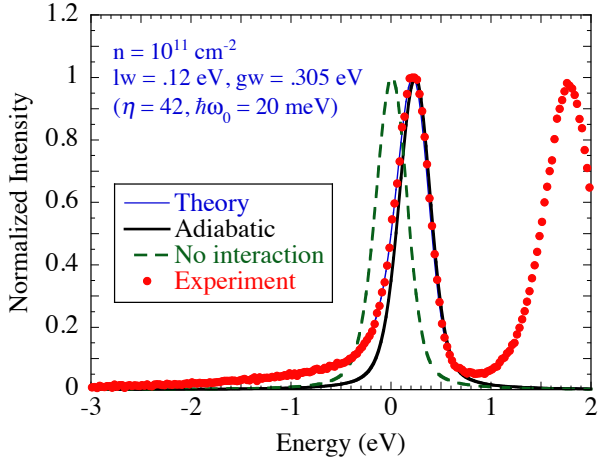}
\caption{(Color online) C 1s core-level spectra of free-standing graphene. The red circles are the experimental spectrum from quasi-free-standing epitaxial graphene on SiC obtained by hydrogen intercalation\,\cite{Leif}. The thick solid line is the peak with adiabatic assumption, the dashed curve is the peak in absence of interaction with the carriers. The thin solid curve is the result from the sudden approximation where shake up effects from electron-hole pair and plasmon excitations are taken into account. The energy is relative to the position the peak would have if all electron states were frozen. Core-level binding energy increases towards the left in the figure.}
\label{figu4}
\end{figure}
When the graphene sheet is doped the dielectric function becomes much more complicated. However it has been derived by several groups\cite{Ser1,Wun,Sarma}. 
The dielectric function in a general point in the complex frequency plane, $W$, away from the real axis is\,\cite{Ser2}
\begin{equation}
\begin{array}{*{20}{l}}
{\varepsilon \left( {Q,W} \right) = \kappa  + \frac{{2\pi {e^2}}}{Q}{D_0}\left\{ {1 + \frac{{{Q^2}}}{{4\sqrt {{Q^2} - {W^2}} }}\left[ {\pi  - f\left( {Q,W} \right)} \right]} \right\};}\\
{f\left( {Q,W} \right) = {\rm{asin}}\left( {\frac{{1 - W}}{Q}} \right) + {\rm{asin}}\left( {\frac{{1 + W}}{Q}} \right)}\\
{\qquad \qquad  - \frac{{W - 1}}{Q}\sqrt {1 - {{\left( {\frac{{W - 1}}{Q}} \right)}^2}}  + \frac{{W + 1}}{Q}\sqrt {1 - {{\left( {\frac{{W + 1}}{Q}} \right)}^2}} ,}
\end{array}
\label{equ46}
\end{equation}
where ${D_0} = \sqrt {4n/\pi {\hbar ^2}{v^2}} $ is the density of states at the Fermi level and $n$ is the doping concentration. In both Eqs. (\ref{equ45}) and (\ref{equ46}) we let $W$ be real valued but add a small imaginary part to get the retarded forms of the dielectric functions.

In Figs.\,\ref{figu1} and \ref{figu2} we show the contour plot and surface plot, respectively, of $-{\mathop{\rm Im}\nolimits} \left[ {\varepsilon {{\left({Q,W} \right)}^{ - 1}}} \right]$ for doped graphene.

The integrands in Eqs.\,(\ref{equ41}), (\ref{equ42}), and (\ref{equ43}) all have an extra factor $1/W$ for small $W$ which means that small energy transfers are important in the shake up structure of the spectra. In Fig.\,\ref{figu3} we give the corresponding surface plot for pristine graphene. Here the excitations are of electron-hole pair type.

We have taken the Lorentzian broadening into account but how should we include the Gaussian instrumental broadening? Here we use a trick.
If all shake up processes were involving only one discrete frequency, ${{\omega _0}}$, the $a$-, $b$-, and $D$-functions would become
\begin{equation}
\begin{array}{l}
{a_g}\left( T \right) = \eta \left[ {1 - \cos \left( {{W_0}T} \right)} \right],\\
{b_g}\left( T \right) =  - \eta \sin \left( {{W_0}T} \right),\\
{D_g} = \eta {W_0},\\
\eta  = \int {\frac{{{d^2}q}}{{{{\left( {2\pi } \right)}^2}}}\frac{{{{\left| {g\left( {\bf{q}} \right)} \right|}^2}}}{{{{\left( {\hbar {\omega _0}} \right)}^2}}}}, 
\end{array}
\label{equ47}
\end{equation}
\begin{figure}
\includegraphics[width=8cm]{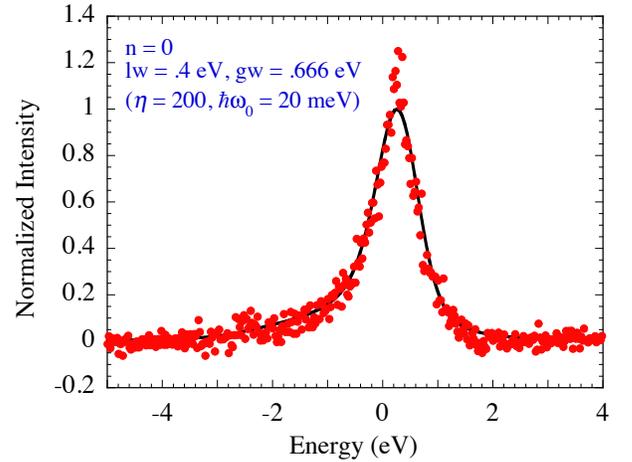}
\caption{(Color online) Experimental ($\mu$-XPS) C 1s core-level spectrum of a suspended graphene sheet obtained in Ref.\,\cite{Wu} using the photon energy 480 eV, red circles. The solid curve is the theoretical result for an undoped free-standing graphene sheet. A linear background was subtracted from the experimental data. We used Lorentzian and Gaussian broadening of .4 eV and .666 eV, respectively}
\label{figu5}
\end{figure}
where $\eta $ is a strength constant and the spectrum would consist of a series of delta functions,
\begin{equation}
\begin{array}{*{20}{l}}
\begin{array}{l}
S\left( W \right) = \frac{1}{{2\pi }}\int\limits_{ - \infty }^\infty  {dT{e^{ - \eta }}{e^{ - i\left( {W - \eta {W_0}} \right)}}\underbrace {{e^{\eta {e^{i{W_0}T}}}}}_{\sum\limits_n {\frac{{{{\left( {\eta {e^{i{W_0}T}}} \right)}^n}}}{{n!}}} }} \\
 = {e^{ - \eta }}\sum\limits_n {\frac{{{\eta ^n}}}{{n!}}\frac{1}{{2\pi }}\underbrace {\int\limits_{ - \infty }^\infty  {dT{e^{ - i\left( {W - \eta {W_0} + n{W_0}} \right)T}}} }_{2\pi \delta \left( {W - \eta {W_0} + n{W_0}} \right)}} 
\end{array}\\
{ = \sum\limits_{n = 0}^\infty  {\underbrace {\frac{{{e^{ - \eta }}{\eta ^n}}}{{n!}}}_{\scriptstyle{\rm{Poisson}}\hfill\atop
\scriptstyle{\rm{distribution}}\hfill}} \delta \left( {W - \eta {W_0} + n{W_0}} \right).}
\end{array}
\label{equ48}
\end{equation}
\begin{figure}
\includegraphics[width=8cm]{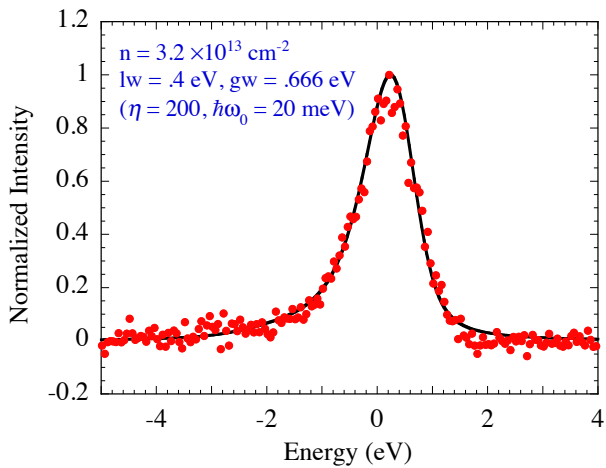}
\caption{(Color online) Same as in Fig.\,\ref{figu5} but now for a gold supported graphene sheet. The theoretical curve was obtained assuming a doping density of $3.2 \times 10^{13} \rm {cm}^{-2}$. We used the same Lorentzian and Gaussian broadening as in Fig.\,\ref{figu5}.}
\label{figu6}
\end{figure}
The amplitudes of the delta functions form a poisson distribution. Why is it a poisson distribution? The average number of bosons surrounding the core hole is $\eta $. The bosons do not interact and the core hole does not recoil when a boson is excited. Thus the probability that a boson is excited in a given instant does not depend on how many bosons are already excited. Then, according to probability theory the probability for having exactly $n$ bosons excited at a certain time is given by the poisson distribution. 

When the strength parameter $\eta $ is large this turns into a Gaussian distribution. It is close to a Gaussian already at  $\eta =5$. The Gaussian FWHM is given by $gw = 2\hbar {\omega _0}\sqrt {\eta 2\ln 2} $.
The trick is now to choose a small enough value for ${\omega _0}$ so that $\eta $ becomes large enough. Then we add the resulting $a$-, $b$-, and $D$-functions to the original functions in Eq.\,(\ref{equ40}). Thus we get the final spectrum with both Lorentzian and Gaussian broadened structures from
\begin{equation}
\begin{array}{l}
S(W) = \frac{1}{\pi }\int\limits_0^\infty  {dT\left[ {{e^{\left( { - \frac{{LW}}{2}T} \right)}}{e^{ - \left[ {a(T) + {a_g}(T)} \right]}}} \right.} \\
\left. {\quad \quad \quad \quad  \times \cos \left\{ {\left[ {W - \left( {D + {D_g}} \right)} \right]T - \left[ {b\left( T \right) + {b_g}\left( T \right)} \right]} \right\}} \right].
\end{array}
\label{equ49}
\end{equation}

This is the relation we have used in finding the XPS spectra presented in Figs.\,\ref{figu4}-\ref{figu7}.

\section{\label{res}Results}

In Fig.\,\ref{figu4} we compare our results to an experimental XPS spectrum\,\cite{Leif0} represented by red circles. The experiment was performed on a single quasi-free-standing epitaxial graphene layer on SiC obtained by hydrogen intercalation\,\cite{Leif}. This leads to a virtually undoped graphene sheet. The photon energy used was 750 eV. 
The leftmost peak is from graphene C\,1s and the rightmost from the C\,1s core level in the SiC substrate. The dashed curve is the result one would get if there were no interaction between the electrons and the core hole. Another way to express this is to say that the electron wave functions are frozen. The energies are given relative to the position of this peak. The peak is symmetric and has the full broadening but no structure from shake-up effects. The thick solid curve is the adiabatic peak. This is what the spectrum would look like if the excitation were adiabatic, i.e., the excitation were so slow that the electrons were left in a quasiequilibrium state; the core-hole potential was fully screened. This peak is just the dashed peak shifted by the energy $d$. The thin solid curve is the full result from Eq.\,(\ref{equ49}). When we used the dielectric function from pristine graphene the curve emerged a little above the experimental tail. When we added a very small doping level, $10^{11} \rm{cm}^{-2}$, the agreement with experiment was perfect. The reason for that a small amount of doping pushes down the tail a little is that there are some extra contributions at the top of the peak and the normalization then leads to a reduction away from the center of the peak.

In Figs.\,\ref{figu5} and \ref{figu6} we compare our results for pristine and doped free-standing graphene to the experimental results in Ref.\,\cite{Wu}.
The experiments were performed using a photon energy of 480 eV on a suspended single graphene sheet and on a gold supported single graphene sheet. In the first sample the graphene is more or less undoped; in the second gold provides p-doping. We subtracted a linear background from both experimental sets of data and used the same broadening parameters for both spectra. The main peak in the doped sample is broader than in the undoped but this extra broadening comes from different shake-up effects in the two samples. The theoretical spectrum was obtained assuming a doping density of $3.2 \times 10^{13} \rm {cm}^{-2}$. This number $3.2 \times 10^{13}$ should not be taken too seriously. We varied the density in equidistant steps on a logarithmic scale and tried $10^{13} \rm {cm}^{-2}$, $10^{13.5} \rm {cm}^{-2}$
and $10^{14} \rm {cm}^{-2}$. The one in the middle gave the best fit with experiments. 
\begin{figure}
\includegraphics[width=8cm]{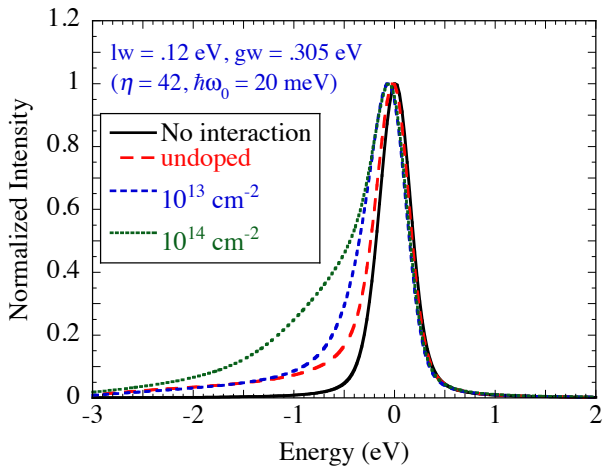}
\caption{(Color online) Shape of the C 1s core-level spectra of free-standing graphene. To get a better view of the peak shapes we have removed the over all shift, $d$, from all spectra.  The symmetric peak is the result from the fully adiabatic approximation and the result when there is no interaction between the core hole and the electron-hole system. The red dashed curve is the result from the undoped graphene sheet, the blue short-dashed from a sheet with doping concentration $10^{13} {\rm cm}^{-2}$, and the green dotted curve from a sheet with doping concentration $10^{14} {\rm cm}^{-2}$, respectively. The curve for doping concentration $10^{12} {\rm cm}^{-2}$ has been omitted since it is very close to the curve for pristine graphene. }
\label{figu7}
\end{figure}

Finally, in Fig.\,\ref{figu7} we show the theoretical results for different doping levels. In order to more clearly see the effect doping has on the peak shape we have here removed all different shifts, $d$, for the curves. The solid black curve is the noninteracting curve or the adiabatic curve (it is the same for all doping levels now when the shift has been removed). The red dashed curve is the pristine result. The blue short-dashed curve is for the doping density $10^{13} {\rm cm}^{-2}$, and the green dotted curve from a sheet with doping concentration $10^{14} {\rm cm}^{-2}$. Important doping effects show up first at doping concentrations exceeding $10^{12} {\rm cm}^{-2}$. The reason is that some electron-hole-pair excitation-channels are blocked with doping due to the Pauli exclusion principle; this is compensated by new electron-hole-pair excitation-channels and the new plasmon-excitation channel. 

In a recent work\,\cite{Desp} one came to the same conclusion as we regarding the need to make a full calculation of the core-hole spectra in graphene instead of using the Doniach and \u{S}unji\'{c} fitting of the peaks. They treated the band-structure of graphene in a different way compared to our and relied on density functional theory. They seem to have assumed a constant 2D plasmon frequency and thereby obtained sharp plasmon structures in the tail region of the peaks. Such structures are not observed in the experimental spectra.

\section{\label{sum}Summary and Conclusions}

To summarize, we have presented a detailed derivation of the core-level spectra of 2D systems and presented numerical results for pristine and doped free-standing graphene. Although, pristine graphene is not a metal its core-level spectrum shows a peak tailing, characteristic of metallic systems. The tailing increases with doping for doping concentrations exceeding  $10^{12} {\rm cm}^{-2}$. The peak shape changes with further increase of the doping concentration. This opens up for a complementary way to estimate the degree of doping of a sample. We have compared our results to three different experimental spectra from two experimental groups. The agreement is quite good which is very encouraging. We have furthermore introduced a convenient way to introduce the effect of Gaussian instrumental broadening in the formalism.

\section{Acknowledgements}
We thank Professor Leif I. Johansson and Professor Chung-Lin Wu for providing us with their experimental data.


\begin{thebibliography}{10}
\bibitem{Leif0}The spectrum was not published before but how the sample was prepared is described in Ref.\,\cite{Leif}.
\end{thebibliography}
\end{document}